\documentclass[sigconf]{acmart}

\usepackage{booktabs} % For formal tables
\usepackage{multirow}
\usepackage{color}
\usepackage{amsmath}
\usepackage{balance} %zj

\AtBeginDocument{%
  \providecommand\BibTeX{{%
    \normalfont B\kern-0.5em{\scshape i\kern-0.25em b}\kern-0.8em\TeX}}}

\copyrightyear{2022}
\acmYear{2022}
\setcopyright{acmcopyright}\acmConference[SIGIR '22]{Proceedings of the 45th International ACM SIGIR Conference on Research and Development in Information Retrieval}{July 11--15, 2022}{Madrid, Spain}
\acmBooktitle{Proceedings of the 45th International ACM SIGIR Conference on Research and Development in Information Retrieval (SIGIR '22), July 11--15, 2022, Madrid, Spain}
\acmPrice{15.00}
\acmDOI{10.1145/3477495.3532053}
\acmISBN{978-1-4503-8732-3/22/07}

\begin{document}
\fancyhead{}

\title{Re-weighting Negative Samples for Model-Agnostic Matching}

\author{Jiazhen Lou}
\authornote{Both authors contributed equally to this paper.}
\affiliation{%
  \institution{Alibaba Group}
  \city{Hangzhou, China}
}
\email{jane.ljz@alibaba-inc.com}

\author{Hong Wen}
\authornotemark[1]
\affiliation{%
  \institution{Alibaba Group}
  \city{Hangzhou, China}
}
\email{qinggan.wh@alibaba-inc.com}

\author{Fuyu Lv}
\affiliation{%
  \institution{Alibaba Group}
  \city{Hangzhou, China}
}
\email{fuyu.lfy@alibaba-inc.com}

\author{Jing Zhang}
\affiliation{%
  \institution{The University of Sydney}
  \city{Darlington NSW 2008, Australia}
}
\email{jing.zhang1@sydney.edu.au}

\author{Tengfei Yuan}
\affiliation{%
  \institution{Alibaba Group}
  \city{Hangzhou, China}
}
\email{yufei.ytf@alibaba-inc.com}

\author{Zhao Li}
\affiliation{%
  \institution{Zhejiang University}
  \city{Hangzhou, China}
}
\email{zhao_li@zju.edu.cn}

\begin{abstract}

Recommender Systems (RS), as an efficient tool to discover users' interested items from a very large corpus, has attracted more and more attention from academia and industry. As the initial stage of RS, large-scale matching is fundamental yet challenging. A typical recipe is to learn user and item representations with a two-tower architecture and then calculate the similarity score between both representation vectors, which however still struggles in how to properly deal with negative samples. In this paper, we find that the common practice that randomly sampling negative samples from the entire space and treating them equally is not an optimal choice, since the negative samples from different sub-spaces at different stages have different importance to a matching model. To address this issue, we propose a novel method named Unbiased Model-Agnostic Matching Approach (UMA$^2$). It consists of two basic modules including 1) General Matching Model (GMM), which is model-agnostic and can be implemented as any embedding-based two-tower models; and 2) Negative Samples Debias Network (NSDN), which discriminates negative samples by borrowing the idea of Inverse Propensity Weighting (IPW) and re-weighs the loss in GMM. UMA$^2$ seamlessly integrates these two modules in an end-to-end multi-task learning framework. Extensive experiments on both real-world offline dataset and online A/B test demonstrate its superiority over state-of-the-art methods.

\end{abstract}




\maketitle

\setlength{\abovecaptionskip}{0.2cm}
\setlength{\belowcaptionskip}{-0.185cm}

\begin{figure}[h]
  \centering
  \includegraphics[width=1\linewidth]{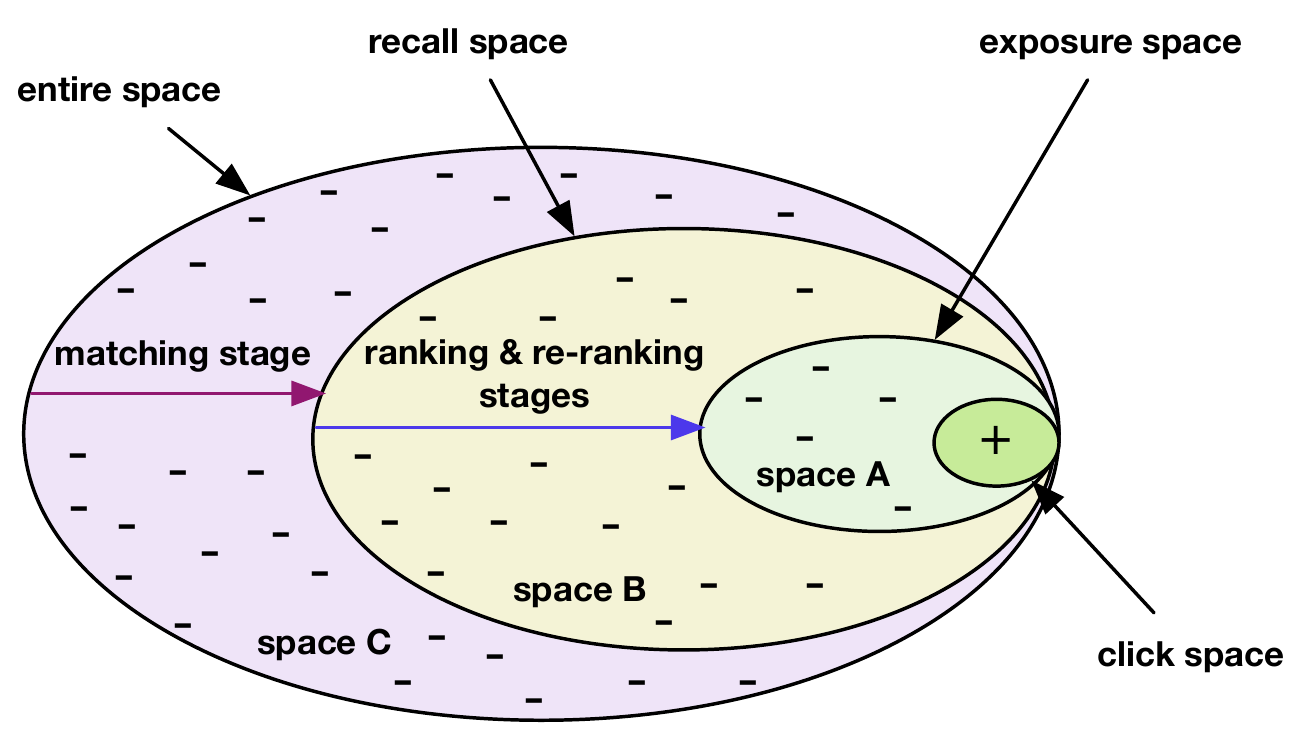}
  \caption{Illustration of the multi-stage pipeline in large-scale recommender systems, where positive samples and negative samples from different sub-spaces are denoted by `+' and `-', respectively. 
  }
  \label{fig:recall}
\end{figure}

\section{Introduction}
Recommender Systems (RS) \cite{zhang2019deep,huseynov2020intelligent,shen2022deep}, which aim at providing personalized items from a very large corpus to users, is of great practical significance in improving users' experience and increasing business revenue~\cite{zhang2020empowering}. A large-scale industrial RS usually responds users’ online requests through a multi-stage pipeline of ``matching$\to$ranking$\to$re-ranking''~\cite{covington2016deep, shen2021sar}, where the ranking stage would take as input results selected from the preceding matching stage. The selection mechanism gradually delivers items from entire space to recall space, then to exposure space as illustrated in Figure~\ref{fig:recall}. 
Only items in exposure space can be displayed to users, of which the items clicked by users are labeled as positive samples. Obviously, constructing a superior matching model for the matching stage is very essential since it determines the quality of candidate item set delivered to follow-up stages, which is the goal of this study.

One straightforward strategy is following the ways of building ranking models~\cite{zhou2018deep, ma2018entire, wen2020entire} to construct a matching model. However, it has obvious shortcomings since they only employ observed data in exposure space, which occupies only a very small portion of entire space, resulting in the dramatically discrepancy
between training space and inference space. This problem is also called Sample Selection Bias (SSB) issue~\cite{chen2020bias}. Though, ESMM~\cite{ma2018entire}, ESM$^2$~\cite{wen2020entire}, and HM$^3$~\cite{wen2021hierarchically} are further proposed to alleviate the SSB issue from the perspective of exposure space modelling to make the post-click conversion rate (CVR) estimation, they focus on employing post-click information that are still limited in exposure space. To go a step forward, ESAM~\cite{chen2020esam} transfers the knowledge from displayed items in exposure space to non-displayed items in recall space for alleviating the distribution inconsistency. But it still neglects the remaining large number of samples in entire space, which still biases the matching model in inference stage. As a result, the SSB issue heavily limits the performance improvement of matching, especially for deep matching models~\cite{huang2013learning,covington2016deep,lv2019sdm,li2019multi,yu2021dual}, which follow a popular two-tower paradigm that learns vector representations of items and users to enable fast k-nearest neighbor retrieval and have been increasingly important in industrial systems.

To resolve the SSB issue, a practical recipe in matching models (YouTube DNN~\cite{covington2016deep}, SDM~\cite{lv2019sdm}, MIND~\cite{li2019multi}, DAT~\cite{yu2021dual}) is randomly sampling negatives from the entire space to ensure the consistency between training and inference phases. 
Moreover, ~\citeauthor{fei2020sample}~\cite{fei2020sample} proposes to mix up observed data and random negatives for sample optimization.
Those work treat all negative samples equally in the final loss, although proving effective, this strategy is not an optimal choice. Referring to Figure~\ref{fig:recall}, negative samples in entire space can be categorized into three disjoint spaces, $i.e.$, entire but un-recalled, recalled but unexposed, and exposed but unclicked, denoted as \textit{Space C}, \textit{B}, \textit{A}, respectively. 
Negative samples in different spaces should be discriminated.
For instance, negative samples in \textit{Space A} are not those that users particularly dislike compared with negative samples in the remaining part of \textit{Space C} and \textit{B}~\cite{ding2019reinforced,lv2020xdm}.

Considering the shortcomings of existing work, we believe it is essential to develop an universal debiasing deep matching model, which can properly utilize entire space samples.
To go a step forward, we propose a novel method named Unbiased
Model-Agnostic Matching Approach (UMA$^2$). 
It is built upon the Multi-Task Learning (MTL) framework and consists of two key modules including a General Matching Model (GMM) and a Negative Samples Debias Network (NSDN). 
GMM that aims to learn user and item representations is model-agnostic and can be implemented with any embedding-based two-tower model. 
NSDN discriminates negative samples from the three disjoint spaces by using Inverse Propensity Weighting (IPW)~\cite{zhang2020large, saito2020doubly}. 
Specifically, it employs two auxiliary tasks namely ``Entire$\to$Recall'' and ``Recall$\to$Exposure'', where the probability from the entire space to the recall space as well as the probability from the recall space to the exposure space are predicted, to implement IPW and reweigh the loss in GMM. 
The contributions of this paper is three-fold:

\begin{itemize}

    \item We propose a novel method named Unbiased Model-Agnostic Matching Approach (UMA$^2$) for the matching stage in RS, which is compatible and can be seamlessly integrated with a variety of two-tower models for learning user and item representations in an end-to-end manner.

    \item Our proposed method can effectively mitigate the SSB issue by discriminating negative samples in entire space from the perspective of IPW, which can be implemented efficiently in a multi-task learning framework.

    \item We conduct extensive experiments on real-world offline dataset and online A/B test. Both results demonstrate the superiority of the proposed UMA$^2$ over representative methods. Now, UMA$^2$ serves millions of users on our platform, achieving 3.35\% CTR improvement.

\end{itemize}

\section{Related Work}
\label{sec:RELATED_WORK}

\textbf{Bias and Debias Recommendation:} Sample Selection Bias (SSB) is a widely-recognized issue due to the dramatically discrepancy between training space and inference space. For example, when training a conversion rate (CVR) model, traditional methods \cite{wen2019multi, lee2012estimating} always employs clicked samples as training set while making predictions on all exposure samples. To alleviate the issue, ESMM \cite{ma2018entire}, ESM$^2$ \cite{wen2020entire} and HM$^3$ \cite{wen2021hierarchically} are from the perspective of exposure space modelling to make the CVR prediction. Another way is resorting to the Inverse Propensity Weighting (IPW) method, which is theoretically unbiased \cite{zhang2020large, saito2020doubly}. In this paper, we also employ the idea of IPW to discriminate negative samples in entire space.

\textbf{Two-Tower Matching Models:} Recently, deep neural networks with two-tower architecture have become the mainstream trend to efficiently retrieve candidates in industry, which firstly learn user and item representations, followed by calculating the similarity score between both representations, $e.g.$, DSSM \cite{huang2013learning}, YouTube DNN \cite{covington2016deep} and DAT \cite{yu2021dual}. 
However, all the methods neglect the side effect of equally treating negative samples. Our work also contributes to the line of research while employing the idea of IPW to discriminate negative samples. Moreover, it is a model-agnostic approach and is compatible with a variety of two-tower models, $e.g.$, YouTube DNN \cite{covington2016deep}, DAT \cite{yu2021dual}, and other related methods \cite{chen2020esam}.

\begin{figure}
  \centering
  \includegraphics[width=1\linewidth]{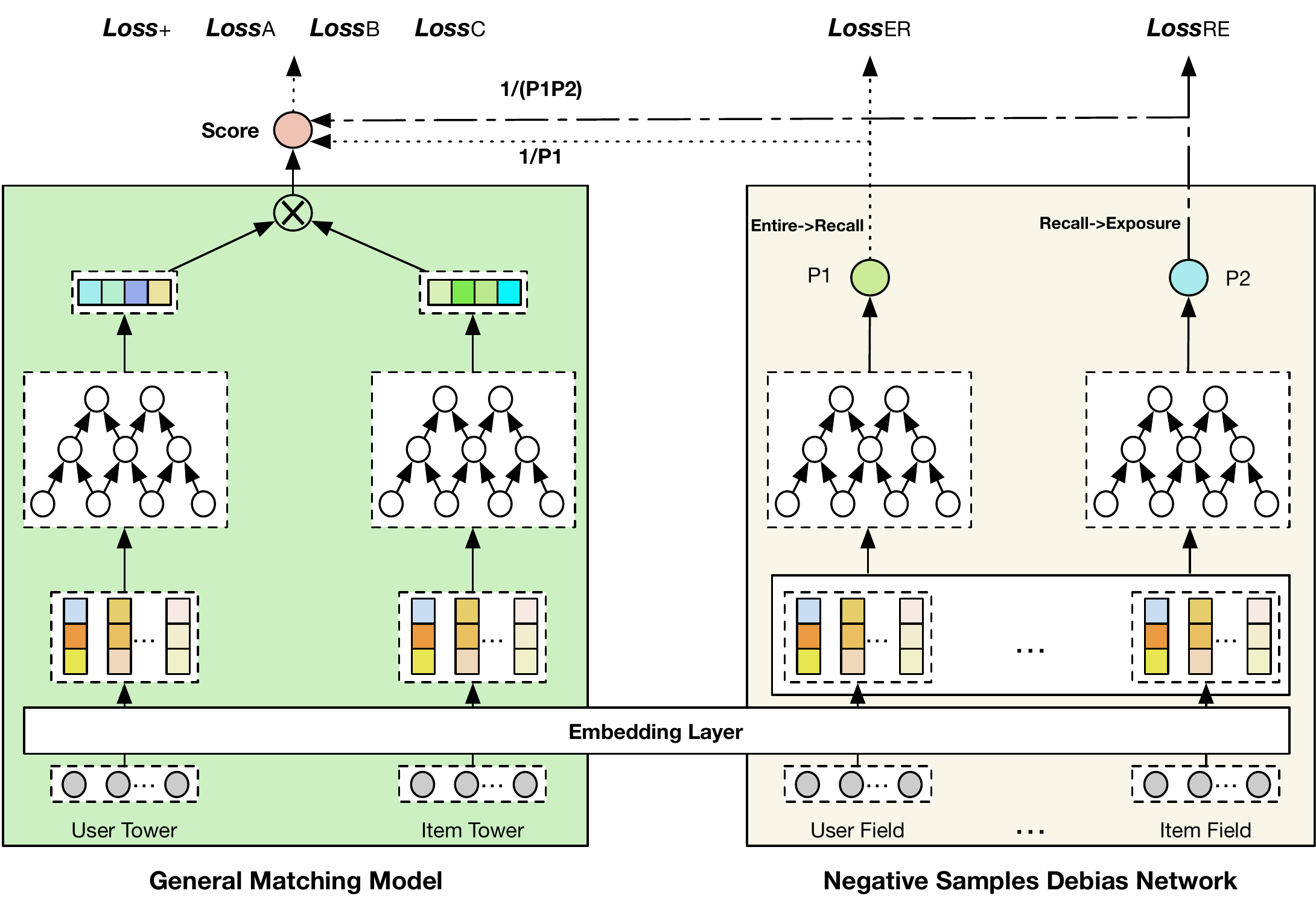}
  \caption{ 
    The architecture of our proposed UMA$^2$.
  }
  \label{fig:two_tower}
\end{figure}

\section{The Proposed Approach}
\label{sec:method}

In this paper, we propose a novel method named Unbiased Model-Agnostic Matching Approach (UMA$^2$), depicted in Figure~\ref{fig:two_tower}. It consists of two basic modules namely General Matching Model (GMM) and Negative Samples Debias Network (NSDN).

Given $\mathcal{U}=\left \{ u_{1},u_{2},...,u_{N} \right \}$, $\mathcal{I}=\left \{ i_{1},i_{2},...,i_{M} \right \}$, $\mathcal{R}=\mathcal{U}\times \mathcal{I}$ be a set of users, items, user-item feedback matrix, respectively, where $N$, $M$ represents the number of users, items, respectively. $R_{u,i}=1$ denotes user $u$ has given a positive feedback to item $i$, $e.g.$, $u$ has clicked $i$, otherwise $R_{u,i}=0$. Each element from $\mathcal{U}$ or $\mathcal{I}$ can be regarded as the concatenation of various types of low dimensional dense representation by transforming original one-hot vector with corresponding embedding matrix. Our learning goal is to deliver a matching model to efficiently retrieve possibly users' interested items from a large-scale item corpus, given a certain user.

\subsection{General Matching Model}
\label{subsec:two_tower_retrieval_model}

The General Matching Model (GMM) module is model-agnostic and can be implemented with any embedding-based two-tower model. In this paper, we resort to the architecture of YouTube DNN \cite{covington2016deep} as the baseline implementation. As illustrated in Figure~\ref{fig:two_tower}, the GMM module consists of user tower and item tower, where the inputs $f_{u}$ (resp. $f_{i}$) of user tower (resp. item tower) contains various types of features from user field (resp. item field), $e.g.$, \emph{users' profiles} and \emph{users' historical behaviors} for user field, \emph{item id} and \emph{item's accumulated CTR} for item field. The goal of the two-tower model is mapping $f_{u}$ (resp. $f_{i}$) into user representation (resp. item representation) via corresponding mapping function $F_{u}(.)$ (resp. $F_{i}(.)$), denoted as $V_{u}=F_{u}(f_{u})$, $V_{i}=F_{i}(f_{i})$, respectively. $F_{u}(.)$ or $F_{i}(.)$ is usually implemented as multi-layer perceptron (MLP) with \emph{Relu} as the activation function. Given $V_{u}$ and $V_{i}$, a scoring function $Score(,)$, $e.g.$, inner-product, is employed to calculate the similarity score between both representation vectors, denoted as $s_{u,i}=Score(V_{u},V_{i})$. Now, for supervising the learning of $s_{u,i}$ as well as maintaining the consistency of training space and inference space, a practical recipe is regarding clicked samples in exposure space as positive samples. Meanwhile, for each positive sample, corresponding negative samples are constructed via a random sampling strategy with respective to the entire space.

\subsection{Negative Samples Debias Network}
\label{subsec:negative_sample_debias_network}

To view all the negative samples in entire space at a finer granularity, they can be further categorized into three disjoint spaces, $i.e.$, entire but un-recalled, recalled but unexposed, and exposed but unclicked, denoted as \textit{Space C}, \textit{B}, \emph{A} in Figure~\ref{fig:recall}, respectively. Negative samples in different spaces should be discriminated.
For instance, negatives in \textit{Space A} are not those that users particularly dislike compared with negatives in \textit{Space C} or \textit{B}~\cite{ding2019reinforced,lv2020xdm}. Therefore, we borrow the idea of Inverse Propensity Weighting (IPW)~\cite{zhang2020large, saito2020doubly} from the causal inference area to handle the problem. Specifically, we propose the Negative Samples Debias Network (NSDN) as the embodiment of the IPW method. It consists of two auxiliary tasks namely ``Entire$\to$Recall'' and ``Recall$\to$Exposure'', which aim to predict the probability from the entire space to the recall space as well as the probability from the recall space to the exposure space, respectively.

Moreover, the IPW method is theoretically unbiased by re-weighing the loss of observed samples \cite{zhang2020large, saito2020doubly}. Taking how to debias negative samples in \textit{Space B} with respect to entire space as an example, we detail the process. First, we define the loss $Loss_{GMM}$ of GMM as follows:

\begin{equation}
    Loss_{GMM} = \frac{1}{|\mathcal{D}|}\sum_{(u,i)\in \mathcal{D}}\frac{o_{u,i}e(y_{u,i},\hat{y}_{u,i})}{\hat{p}_{u,i} } 
    = \frac{1}{|\mathcal{O}|}\sum_{(u,i)\in \mathcal{O}}\frac{e(y_{u,i},\hat{y}_{u,i})}{\hat{p}_{u,i} },
\label{eq:ipw_form}
\end{equation}
where $e(y_{u,i}, \hat{y}_{u,i})$ is the cross-entropy loss between real $y_{u,i}$ and predicted $\hat{y}_{u,i}$ in GMM. $\mathcal{D}$ refers to the entire space including all the $(u,i)$ pairs. $o_{u,i}=1$ denotes that item $i$ is recalled for user $u$, $i.e.$, observed data, otherwise $o_{u,i}=0$. Collecting all the observed data from space $\mathcal{D}$ forms the recall space $\mathcal{O}$. Intuitively, the easier item $i$ recalled for user $u$, the smaller sample weight the corresponding user-item pair should has. Here, we define $\hat{p}_{u,i}$ ($i.e.$, $p_{1}$), as the probability of the item $i$ being recalled for user $u$. It can be predicted from an auxiliary task ``Entire$\to$Recall'', where recalled items are positive samples while the items randomly sampled from the \textit{Space C} are negative ones. 

In this way, negative samples in \textit{Space B} are re-weighed by multiplying corresponding weight $1/p_{1}$ in the final loss in GMM. In the same way, we can debias negative samples in \textit{Space A} with respect to recall space by multiplying sample weight $1/p_{2}$, where $p_{2}$ can be predicted from an auxiliary task ``Recall$\to$Exposure''. Furthermore, negatives in \textit{Space A} can be debiased with respect to entire space by multiplying weight $1/p_{1}p_{2}$ in the final loss in GMM.

\subsection{Model Training}
\label{subsec:model_training}

We treat the matching problem as a binary classification problem, and the final loss defined as :
\begin{equation}
    loss=loss_{+}+\lambda _{1}loss_{\mathcal{A}}+\lambda _{2}loss_{\mathcal{B}}+\lambda _{3}loss_{\mathcal{C}}+\lambda _{4}loss_{ER}+\lambda _{5}loss_{RE},
\label{eq:ipw_form}
\end{equation}
where $loss_{+}$ represents the cross-entropy loss of positive samples, $loss_{\mathcal{A}}$ and $loss_{\mathcal{B}}$ denote the cross-entropy loss of negative samples in \textit{Space A} and \textit{Space B} with the sample weight $1/p_{1}p_{2}$ and $1/p_{1}$, respectively. $loss_{\mathcal{C}}$ indicates the cross-entropy loss of negative samples in \textit{Space C} with an equal weight for each sample. $loss_{ER}$ and $loss_{RE}$ represent the cross-entropy loss for the auxiliary task ``Entire$\to$Recall'' and ``Recall$\to$Exposure'', respectively. Meanwhile, we make \textit{Space B} and \textit{Space C} by sampling a certain amount of items from online real recalled item set with respect to each exposure sample, as well as sampling a certain amount of items from entire space with respect to each recalled item, respectively. $\lambda _{1}$, $\lambda _{2}$, $\lambda _{3}$, $\lambda _{4}$, and $\lambda _{5}$ are hyper-parameters to balance the losses.

\begin{table*}[htbp]
\caption{Model performance with different negative sampling strategies.}
\begin{tabular}{c|c|c|cc|cc|cc}
\hline
\multirow{2}{*}{}                 & \multirow{2}{*}{Sampling Strategy}                                  & \multirow{2}{*}{Competitors}             & \multicolumn{2}{c|}{HitRate}                                         & \multicolumn{2}{c|}{Precision}                                       & \multicolumn{2}{c}{Recall}                                          \\ \cline{4-9} 
                                  &                                                                           &                                          & \multicolumn{1}{c|}{@100}             & @200                         & \multicolumn{1}{c|}{@100}             & @200                         & \multicolumn{1}{c|}{@100}             & @200                         \\ \hline
\multirow{12}{*}{Main results}    & \multirow{2}{*}{SS-A}                                                        & YouTube DNN & \multicolumn{1}{c|}{0.35567}          & 0.39642                      & \multicolumn{1}{c|}{0.00385}          & 0.00213                      & \multicolumn{1}{c|}{0.33902}          & 0.37977                                                  \\  
                                  &                                     & DAT                                     & \multicolumn{1}{c|}{0.37644}          & 0.41831                      & \multicolumn{1}{c|}{0.00412}          & 0.00227                      & \multicolumn{1}{c|}{0.36068}          & 0.40255                      \\ \cline{2-9} 
                                  & \multirow{4}{*}{SS-AB}                                                      & ESAM\_YTB   
                                  & \multicolumn{1}{l|}{0.38040}          & \multicolumn{1}{l|}{0.43422} & \multicolumn{1}{l|}{0.00413}          & \multicolumn{1}{l|}{0.00234} & \multicolumn{1}{l|}{0.36355}          & \multicolumn{1}{l}{0.41736} \\ 
                                  &                                                                           & ESAM\_DAT                                & \multicolumn{1}{l|}{0.40928}          & \multicolumn{1}{l|}{0.47505} & \multicolumn{1}{l|}{0.00443}          & \multicolumn{1}{l|}{0.00254} & \multicolumn{1}{l|}{0.38977}          & \multicolumn{1}{l}{0.45554} \\
                                  &                                                                           & UMA$^2$\_YTB                               & \multicolumn{1}{c|}{0.57014}                & 0.60759                            & \multicolumn{1}{c|}{0.00630}                & 0.00334                            & \multicolumn{1}{c|}{0.54986}                & 0.58731                            \\ 
                                  &                                                                           & UMA$^2$\_DAT                               & \multicolumn{1}{c|}{0.58481}                & 0.62066                            & \multicolumn{1}{c|}{0.00647}                & 0.00342                            & \multicolumn{1}{c|}{0.56513}                & 0.60098                            \\ \cline{2-9} 
                                  & \multirow{4}{*}{\begin{tabular}[c]{@{}c@{}}SS-ABC(random)\end{tabular}} & FM                                       & \multicolumn{1}{l|}{0.58640}          & \multicolumn{1}{l|}{0.69417} & \multicolumn{1}{l|}{0.00816}          & \multicolumn{1}{l|}{0.00513} & \multicolumn{1}{l|}{0.49372}          & \multicolumn{1}{l}{0.60717} \\
                                  &                                                                           & YouTube DNN                              & \multicolumn{1}{c|}{0.73796}          & 0.79481                      & \multicolumn{1}{c|}{0.00817}          & 0.00437                      & \multicolumn{1}{c|}{0.72038}          & 0.77723                      \\
                                  &                                                                           & MIND                                     & \multicolumn{1}{l|}{0.75550}          & \multicolumn{1}{l|}{0.80771} & \multicolumn{1}{l|}{0.00838}          & \multicolumn{1}{l|}{0.00445} & \multicolumn{1}{l|}{0.73897}          & \multicolumn{1}{l}{0.79119} \\
                                  &                                                                           & DAT                                      & \multicolumn{1}{c|}{0.76318}          & 0.81378                      & \multicolumn{1}{c|}{0.00846}          & 0.00448                      & \multicolumn{1}{c|}{0.74629}          & 0.79690                      \\ \cline{2-9} 
                                  & \multirow{2}{*}{\begin{tabular}[c]{@{}c@{}}SS-ABC(fixed)\end{tabular}} & UMA$^2$\_YTB                               & \multicolumn{1}{c|}{0.74693} 
                                  & 0.80140             & \multicolumn{1}{c|}{0.00827} 
                                  & 0.00441             & \multicolumn{1}{c|}{0.72931} 
                                  & 0.78379             \\
                                  &                                                                           & UMA$^2$\_DAT                               & \multicolumn{1}{c|}{\textbf{0.76679}} & \textbf{0.81654}             & \multicolumn{1}{c|}{\textbf{0.00850}} & \textbf{0.00450}             & \multicolumn{1}{c|}{\textbf{0.74994}} & \textbf{0.79969}             \\ \hline
\multirow{2}{*}{Ablation Study 1} & \multirow{2}{*}{\begin{tabular}[c]{@{}c@{}}SS-ABC(fixed)\end{tabular}} & YouTube DNN   & \multicolumn{1}{l|}{0.64448}          & \multicolumn{1}{l|}{0.70377} & \multicolumn{1}{l|}{0.00714}          & \multicolumn{1}{l|}{0.00386} & \multicolumn{1}{l|}{0.62559}          & \multicolumn{1}{l}{0.68488} \\               
 
                                  &                                                                           & DAT                          & \multicolumn{1}{l|}{0.65382}          & \multicolumn{1}{l|}{0.71174} & \multicolumn{1}{l|}{0.00724}          & \multicolumn{1}{l|}{0.00391} & \multicolumn{1}{l|}{0.63476}          & \multicolumn{1}{l}{0.69268} \\ \hline
\multirow{2}{*}{Ablation Study 2} & \multirow{2}{*}{\begin{tabular}[c]{@{}c@{}}SS-ABC(fixed)\end{tabular}} & UMA$^2$\_SOS\_YTB   & \multicolumn{1}{l|}{0.64737}          & \multicolumn{1}{l|}{0.71090} & \multicolumn{1}{l|}{0.00714}          & \multicolumn{1}{l|}{0.00389} & \multicolumn{1}{l|}{0.62791}          & \multicolumn{1}{l}{0.69144} \\               
 
                                  &                                                                           & UMA$^2$\_SOS\_DAT                          & \multicolumn{1}{l|}{0.66201}          & \multicolumn{1}{l|}{0.72552} & \multicolumn{1}{l|}{0.00731}          & \multicolumn{1}{l|}{0.00397} & \multicolumn{1}{l|}{0.64239}          & \multicolumn{1}{l}{0.70590} \\ \hline
\end{tabular}
\label{tab:all_method}
\end{table*}

\section{Experiments}
\label{sec:experiment}

\subsection{Experiment settings}
\label{subsec:settings}

\subsubsection{Dataset preparation}
\label{subsubsec:dataset}
To the extent of our knowledge, there are no public datasets suited for the proposed model since they lack real online recalled items with respect to each exposure sample. To fill this gap, we make the offline dataset by collecting users' real traffic logs from our platform in 30 days, $i.e.$, from 2021-12-11 to 2022-01-09. The offline dataset are divided into the disjoint training set and testing set, where the training set is from 2021-12-11 to 2022-01-08, while the left is the testing set, as summarized in Table~\ref{tab:dataset}.

\begin{table}[htbp]
  \caption{Statistics of the offline dataset.}
  \label{tab:dataset}
  \centering
    \begin{tabular}{c|cc}
      \hline
      Description  & Training Set  & Testing Set \\
      \hline
      \#users &1,629,473 &93,473 \\
      \#items &282,728 &261,914 \\
      \#samples &24,547,397 &1,033,742 \\
      \#positive samples &948,560 &39,779 \\
    
      \hline
  \end{tabular}
\end{table}

\subsubsection{Competitors}
\label{subsubsec:competitors} We compare the proposed method UMA$^2$ with following methods: 

\begin{itemize}
    \item \textbf{FM} \cite{rendle2010factorization}: It can model interactions between users' and items' variables using factorized parameters.
    \item \textbf{YouTube DNN} \cite{covington2016deep}: It is one of the most successful two-tower separate architecture used for matching stage.
    \item \textbf{MIND} \cite{li2019multi}: It captures users' diverse interests with multiple representation vectors.
    \item \textbf{DAT} \cite{yu2021dual}: It customizes an augmented vector for each user-item pair to mitigate the lack of information interaction between both towers.
    \item \textbf{UMA$^2$\_DAT} (resp. \textbf{UMA$^2$\_YTB}): It employs DAT \cite{yu2021dual} (resp. YouTube DNN \cite{covington2016deep}) as the implementation of GMM in UMA$^2$.
    \item \textbf{ESAM\_DAT} (resp. \textbf{ESAM\_YTB}): It employs DAT \cite{yu2021dual} (resp. YouTube DNN \cite{covington2016deep}) as the implementation of the model for source domain in ESAM~\cite{chen2020esam}.
    \item \textbf{UMA$^2$\_SOS\_YTB} (resp. \textbf{UMA$^2$\_SOS\_DAT}): It replaces NSDN by the Sample Optimization Strategy (SOS) proposed in \cite{fei2020sample}, where YouTube DNN \cite{covington2016deep} (resp. DAT \cite{yu2021dual}) is employed as the implementation of GMM.
\end{itemize}

\subsubsection{Evaluation Metrics}
\label{subsubsec:metrics}
To evaluate the effectiveness of all the competitors, we employ widely used metrics \cite{lv2019sdm, yu2021dual} for matching evaluation, $i.e.$, HitRate@K, Precision@K and Recall@K, where \emph{K} is set as a relatively large number, $e.g.$, 100 or 200, since the goal of matching is required to deliver a proper subset to follow-up phases.

\subsubsection{Implementation details}
\label{subsubsec:Implementation_details}

To make the offline evaluation fair, confident, and comparable, all the competitors employ same number of positive and negative samples, share same input features of users and items, and are implemented by distributed Tensorflow 1.4, where learning rate, mini-batch, optimizer, set as 0.001, 512, Adam, respectively. In addition, the dimension of the final user and item's representation vectors are both set to 32, the number of MLP layers in each tower are 4, with dimensions 512, 256, 128 and 32, respectively. Moreover, we implement four kinds of sampling strategies including: 1) \textbf{SS-A}: random negative samples from \textit{Space A}; 2) \textbf{SS-AB}: mixing up negative samples in \textit{Space A} and random negative samples from \textit{Space B}; 3) \textbf{SS-ABC(random)}: random negative samples from entire space; 4) \textbf{SS-ABC(fixed)}: random negative samples from \textit{Space A}, \emph{B} and \emph{C} at a ratio of 1:4:20. 

\subsection{Main results}
\label{subsubsec:results}

\subsubsection{Model Comparison}
\label{subsubsec:Performance_Comparison}
Based on the results summarized in Table~\ref{tab:all_method}, we have following observations: 

\begin{itemize}
    \item[\textbf{1)}] Compared with the implementation methods in \emph{SS-A} strategy such as YouTube DNN or DAT, competitors implemented in other strategies consistently achieve extremely significant improvements, $e.g.$, the proposed UMA$^2$\_YTB in \emph{SS-ABC(fixed)} obtains the lift of HitRate@200, Precision@200, Recall@200 by 91.58\%, 94.06\%, 94.70\%, respectively, over YouTube DNN, which demonstrates the SSB issue caused by the \emph{SS-A} strategy indeed seriously degenerates the performance of a matching model. \item[\textbf{2)}] The improvement obtained by the competitors in \emph{SS-AB} strategy over competitors in \emph{SS-A} strategy highlights the advantage of exploring negative samples in larger space, $e.g.$, \textit{Space B}. Meanwhile, we observe that UMA$^2$\_YTB (resp. UMA$^2$\_DAT) has a superior performance over ESAM\_YTB (resp. ESAM\_DAT), which verifies the effectiveness of debiasing negative samples using the proposed method.
    \item[\textbf{3)}] For the competitors implemented in \emph{SS-ABC(random)} strategy, they all achieve better performance than methods implemented in \emph{SS-A} or \emph{SS-AB} strategy, which once again demonstrates the superiority of exploring negative samples from larger space, $e.g.$, \emph{Space C}. Meanwhile, the improvement obtained by YouTube DNN compared with FM highlights the effectiveness of embodying valuable side information from users or items into a deep neural network to learn user and item representations. Considering that users have multiple interests, MIND performs better than YouTube DNN. As for DAT, it can exploit the information interaction between user tower and item tower while being neglected by MIND. Consequently, it achieves an improvement of 0.75\% HitRate@200, 0.72\% Precision@200, 0.72\% Recall@200 over MIND, respectively.
    \item[\textbf{4)}] To go a step forward, we implement two methods namely UMA$^2$\_YTB and UMA$^2$\_DAT in \emph{SS-ABC(fixed)} strategy. We find that compared with YouTube DNN in \emph{SS-ABC(random)} strategy, UMA$^2$\_YTB obtains 0.83\% HitRate@200, 0.92\% Precision@200, 0.84\%, Recall@200 improvement, respectively; and compared with DAT in \emph{SS-ABC(random)} sampling strategy, UMA$^2$\_DAT obtains 0.34\% HitRate@200, 0.38\% Precision@200, 0.35\% Recall@200 improvement, respectively, which also demonstrate the effectiveness of debiaing negative samples using the proposed method.
\end{itemize}

\subsubsection{Ablation Studies}
\label{subsubsec:Ablation_Study}

1) Fixing the number of negative samples in three disjoint spaces, we implement YouTube and DAT. We observe that UMA$^2$\_DAT (resp. UMA$^2$\_YTB) consistently achieves a superior performance over DAT (resp. YouTube DNN); and 2) We regard sample optimization strategy in \cite{fei2020sample} as another debiasing method. Unsurprisingly, our method achieves a better performance. In a nutshell, both results from different perspectives demonstrate the effectiveness of debiaing negative samples using our method.

\subsubsection{Online A/B Test}
\label{subsubsec:online_test}
We also deploy UMA$^2$\_YTB on our platform for A/B test, where baseline model is YouTube DNN. The proposed UMA$^2$\_YTB achieves average $\boldsymbol{3.35\%}$ CTR gain over the baseline model in successive fifteen days, which is consistent with the offline evaluation results, indicating a significant business value.

\section{Conclusion}
\label{sec:Conclusion}

In this paper, we propose a novel matching method named UMA$^2$ to discriminate negative samples effectively, which consists of a GMM and a NSDN. GMM is model-agnostic and can employ any two-tower model to learn user and item representations. NSDN reweighs negative samples by exploring the idea of IPW for re-weighing the loss in GMM. Experiments on offline datasets and online A/B test demonstrate its superiority.

\balance
\bibliographystyle{ACM-Reference-Format}
\bibliography{sample-base}

\end{document}